\documentclass[12pt,article]{article}
\usepackage{fancyhdr}
\usepackage{float}
\usepackage{textcomp}

\usepackage{amsmath,amssymb}
\usepackage{physics}
\usepackage{wrapfig,lipsum}
\usepackage{mathtools}
\usepackage{longtable}
\usepackage{graphicx}
\usepackage{color}

\begin{document}

 %\huge{
 %\small{

 % Don't want date printed
 \date{}
 
 % Make title bold and 14 pt font (Latex default is non-bold, 16pt)
 \title{\Large\bf An investigation of the Brumadinho Dam Break with HEC RAS simulation}
 
 % For single author (just remove % characters)
 %I. M. Anonymous \\

 \author{ 
 Arun Raman\textsuperscript{1}, Fei Liu\textsuperscript{2} \\
 \textsuperscript{1} \textit{New Jersey West Windsor - Plainsboro High School N} \\
 \textsuperscript{2} \textit{New Jersey Science Academy, Fei.Liu@njsci.org}
 }

 \maketitle

 \section*{\centering Abstract}
 
The Brumadinho dam disaster occurred on the 25 January, 2019 when Dam I, an upstream tailings dam at the Corrego do Feijao iron ore mine, 9 kilometres (5.6 mi) east of Brumadinho, Minas Gerais, Brazil, suffered a catastrophic failure ~\cite{DB19}. Over 248 people died and over \$2.88 billion worth of property were lost or damaged ~\cite{BBC19} ~\cite{BBC} due to the subsequent mud flow and flooding. This is merely 4 years after the previous Mariana dam break affecting over 1 million people downstream due to iron ore mining waste flowing into river basin ~\cite{Fern2016}. To prevent a similar tragedy from reoccurring, it's useful to examine the cause of the Brumadinho dam break and compare observations with model simulations. HEC-RAS, developed by US Army Corps of Engineers ~\cite{HR19A} ~\cite{HR19B}, is used to model the mud flow from the Brumadinho dam break based on the NASA SRTM elevation dataset over Brazil ~\cite{SRTM}. The extent of the mud flow from the HEC-RAS simulation matches the actual flooding due to the dam break. This simulation technique can later be used for future dam collapse predictions. 
 
 \textit{}

 \section{Introduction}

 The primary function of a dam is to store water from flowing downstream. Due to the large gravitational potential energy stored in the elevated water, a dam break can lead to catastrophic results for the area downstream. Since the Brumadinho dam is a tailings dam, focus was primarily given on understanding their design. In most mineral extraction processes, large quantities of water are necessary for the extraction of minerals from ore. Tailings describe the remnant rock particles with a grain size distribution from medium sand to clay-sized particles that are usually deposited in slurry form inclusive of chemicals. A tailings dam is a type of dam mainly used in mining operations to store tailings. When tailings dam failures occur, the tailings often tend to liquefy and flow over a long distance (or sometimes travel along rivers), which can lead to both loss of life and environmental damage. Typical dam failure modes include overtopping, slope instability, seepage and piping, foundation failure and earthquake-induced failure. There are three raised tailings dam designs as discussed in \cite{Klohn}: upstream, downstream, and centerline.
 
 Upstream tailings dams are built gradually "upstream" of the starter dam by integrating tailings materials into the dam for support through the deposition tailings.
While using tailings materials to build the dam reduces construction costs, upstream tailings dams may be less stable than other designs under heavy loads and during earthquakes, as tailings materials may experience liquefaction and lose strength.
They are best suited for arid climates where less water is stored in the impoundment, and in aseismic regions. The shortcoming of this design is the primary cause of failure for the Brumadinho dam.

 Downstream or centerline dams are based on alternative designs that provide more stability. Downstream tailings dams are raised gradually "downstream" of the starter dam, with an internal drain or filter. They often have an impervious layer on the upstream slope of the dam. Downstream tailings dams do not require support from the tailings placed against the core, however tailings placed upstream of the core reduce the hydraulic gradient through the core zone. 
Downstream tailings dams require more material to build than upstream tailings dams but are considered more stable than their upstream counterparts in the event of an earthquake.

 Centreline tailings dams are raised gradually while maintaining the original centreline of the starter dam. They usually have an impervious core supported by a localized zone of compacted tailings, and an internal drain.
The use of cyclone sand tailings (coarse fraction of tailings materials) upstream of the core reduces hydraulic gradients through the core.
Centreline tailings dams do not necessitate more material to build than downstream tailings dams and are more stable than upstream dams in the event of an earthquake

Historically, large dam breaks have caused significant damage to both human life and property.

\begin{table}[H]
\centering
\begin{tabular}{lllll}
\hline
\textbf{Date} & \textbf{Location} & \textbf{Release} & \textbf{Impacts} \\ \hline \\
11/21/2015 & San Kat Kuu, Myanmar & Unknown & 113 people killed. \\\\
 9/9/2008 & Taoshi, China & 190,000 $m^3$ & 277 people killed and 33 injured. \\\\
7/10/1985 & Stava, Italy& 200,000 $m^3$ & 268 people killed \\\\ \hline
\end{tabular}
\caption{Historical failures of large dams ~\cite{WISEUranium}}
\end{table}

Recent failures of tailings dams, such as Mount Polley in Canada in 2014 and Samarco's Fundao facility in Brazil in 2015, have attracted the public's attention and brought the safety of tailings dams to the forefront of community concerns. Agreesive pro-development policies that lessen mining licensing requirements and fast-track mineral exploitation lead to severe environmental risk where around 230,000 mining dams are examined and 45 of which could collapse immediately ~\cite{BRAZ2017} ~\cite{Jenn2019}. A new upstream tailings dam failure just occurred in Brazil near Nossa Senhora do Livramento, where tailings from gold mining were stored, causing power failure and loss of telephone services in countless homes.

As a result, dam break study is a budding and active field of research that can help to prevent human life and property damage. Early work by Wahl ~\cite{WAHL09} evaluated the performance of three embankment dam breach models SIMBA developed by USDA-ARS ~\cite{Temple2005}, HR-BREACH at HR Wallingford, Great Britain ~\cite{Mohamed2}, and FIREBIRD BREACH at Montréal Polytechnic~\cite{wang2002}~\cite{wang2006}. The study is intended to provide an evaluation of modelling technologies that can be integrated with state-of-the-art dam failure flood routing and inundation analysis tools.

Recent advancement of GIS (Geographic Information System) technology ~\cite{GIShis} ~\cite{Devantier1993} has enabled dam-break simulations to use topography information more easily with software such as HEC-RAS and Google Earth. BOUSSEKINE et al ~\cite{BOUS16} studied dam break simulation of Hammam Grouz in Algeria, for a 100-years recurrence flood with the dam break occurring near the peak of the flood event. Their main focus is to examine the arrival time of flooding from the dam to the residential area downstream providing early warning for people living downstream. HOOGESTRAAT ~\cite{HOOG11} presents a detailed technical discussion of using HEC-RAS software for modelling dam failures. Some of the approaches such as level pool routing and 2D flow area in this technical report are experimented with for dam-break simulation in this study. 

\section{Description of the event}

\begin{figure}[H]
      \centering
      \includegraphics[width=16cm]{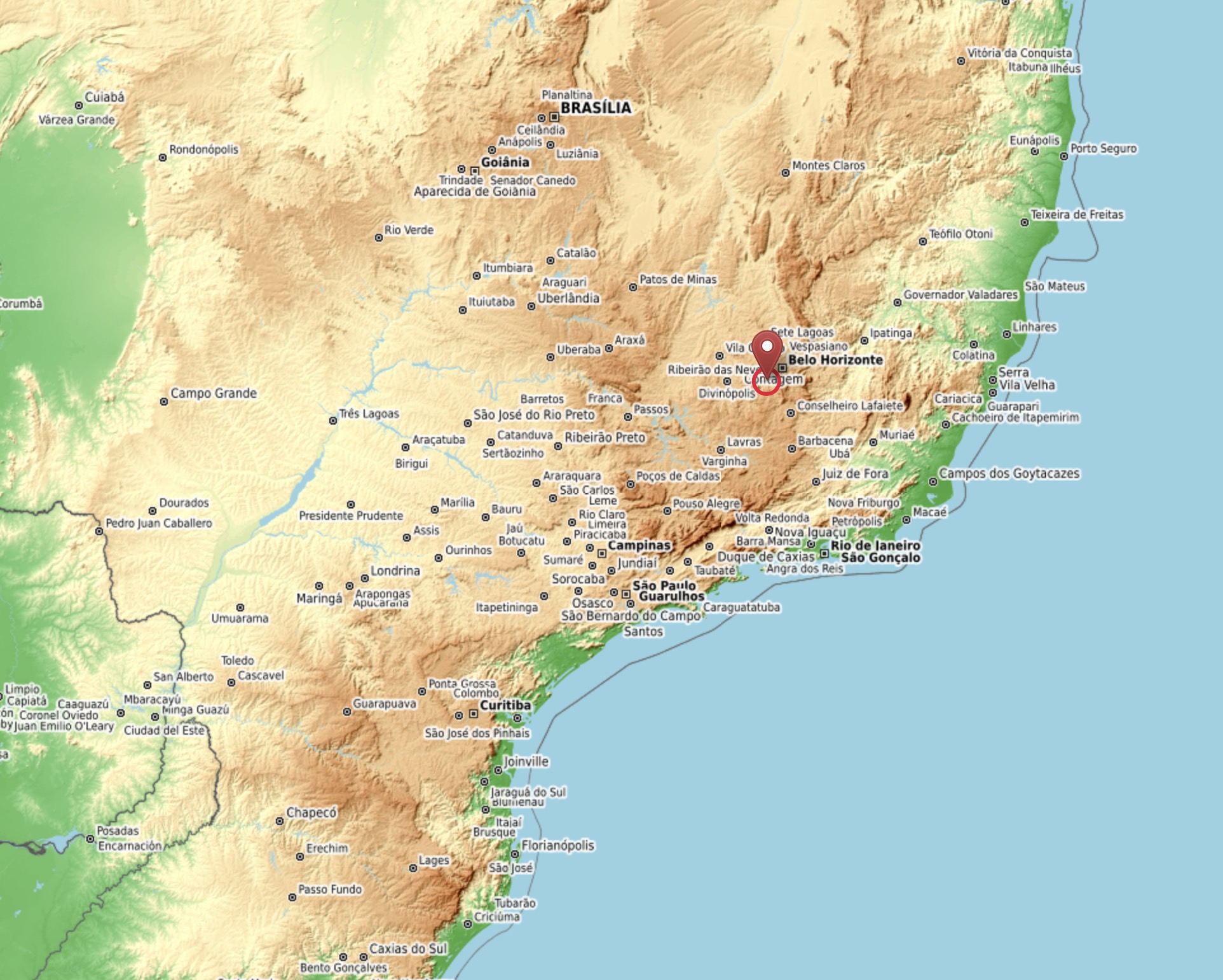}
      \caption{Location of the Brumadinho dam and the city of Brumadinho in Southeastern Brazil}
      \label{fig:geoloc_Brumd}
  \end{figure}

\begin{figure}[H]
      \centering
      \includegraphics[width=16cm]{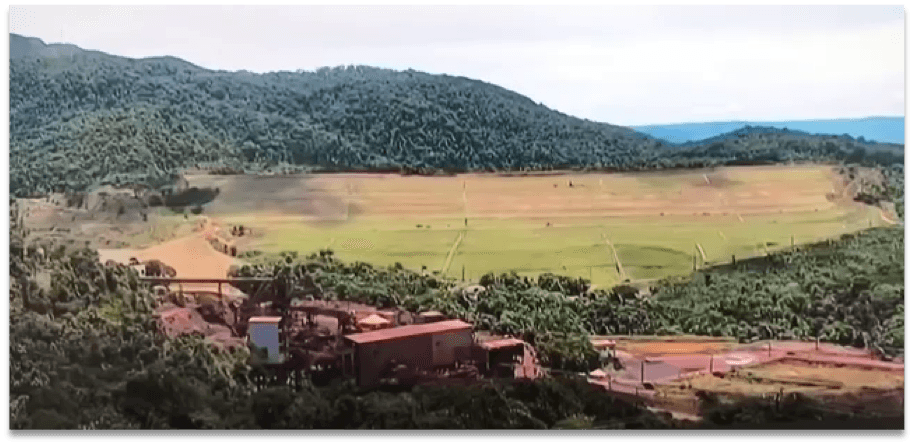}
      \caption{Dam at the moment of the collapse}
      \label{fig:atthemoment}
  \end{figure}

 Brumadinho dam is an upstream tailings dam located in Brumadinho, Brazil as shown in Figure ~\ref{fig:geoloc_Brumd}. The Brumadinho dam was built in 1976 for the storage of iron ore tailings, remnants from the ore's processing ~\cite{DB19}. It lied at an average elevation of 86 meters and held over 12 million cubic meters of tailings ~\cite{DB19} ~\cite{BBC}. The disaster left over 110 dead with only 71 bodies identified. Moreover, 2.7 million square meters of forests, spread out over the 98 km path of the slurry flow, were destroyed ~\cite{DB19}.
It is believed that sliding is the most likely cause of dam failure. From Figure ~\ref{fig:atthemoment}, it is visible that the bottom portion of the dam is slowly creeping forward. This action diminishes the stability of the top portion, causing a total collapse and subsequent flood. The failure was caused by a design error present in upstream tailings dam. Evidence from the video footage suggests that the Brumadinho dam failure can be explained by the analysis presented in ~\cite{Sant2019} and mismanagement of the mining company. Tailings dam requires continual monitoring for potential safety risks as the risk of failure increases with time, often inadequately examined in a laboratory environment. As noted in ~\cite{Sant2019}, an August 2018 (5 months before the failure) audit of the Brumadinho dam yielded the conclusion that the dam was stable. 

\begin{figure}[H]
    
      \centering
      
      \includegraphics[width=16cm]{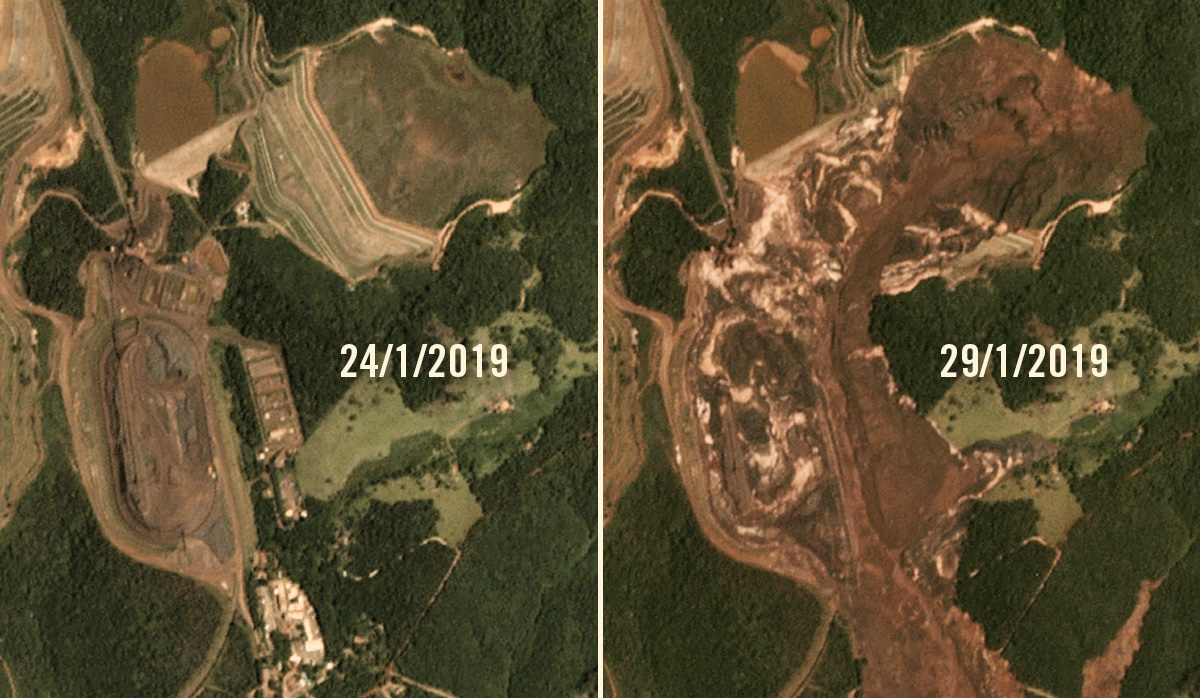}
      
      \caption{Satellite image of the dam before and after the dam break}
    
      \label{fig:before_and_after}
    
  \end{figure}

 The Brumadinho dam broke during local lunchtime and subsequently the mud stored in the dam flew to the Paraopeba river after around 3 hours from the initial break. The event lasted a few more days as mud kept creeping both upstream and downstream along the river. The effect of the dam break near the dam can be seen in Figure ~\ref{fig:before_and_after} ~\cite{Arial2019}. Due to the unfortunate timing of the break, 
most casualties took place in the dam worker's cafeteria as it was lunch hour when the dam broke. 

\begin{figure}[H]
    \begin{minipage}[b]{0.45\linewidth}
      \centering
      
      \includegraphics[width=9cm]{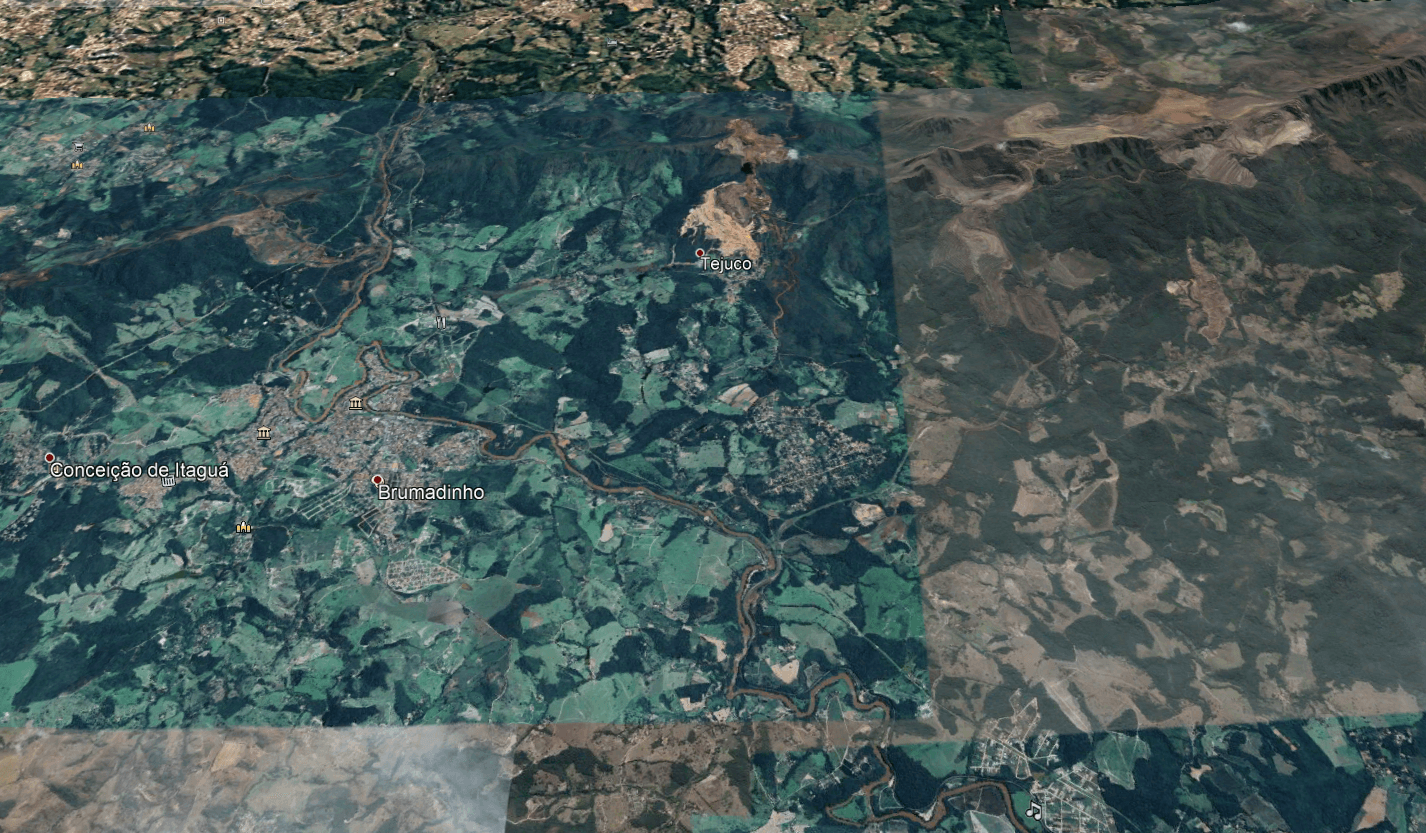}
      
      \caption{The Brumadinho dam in January 2019 before collapse}
      \label{fig:Jan2019B}
    \end{minipage}
    \hspace{1cm}
    \begin{minipage}[b]{0.45\linewidth}
      \centering
      
      \includegraphics[width=9cm]{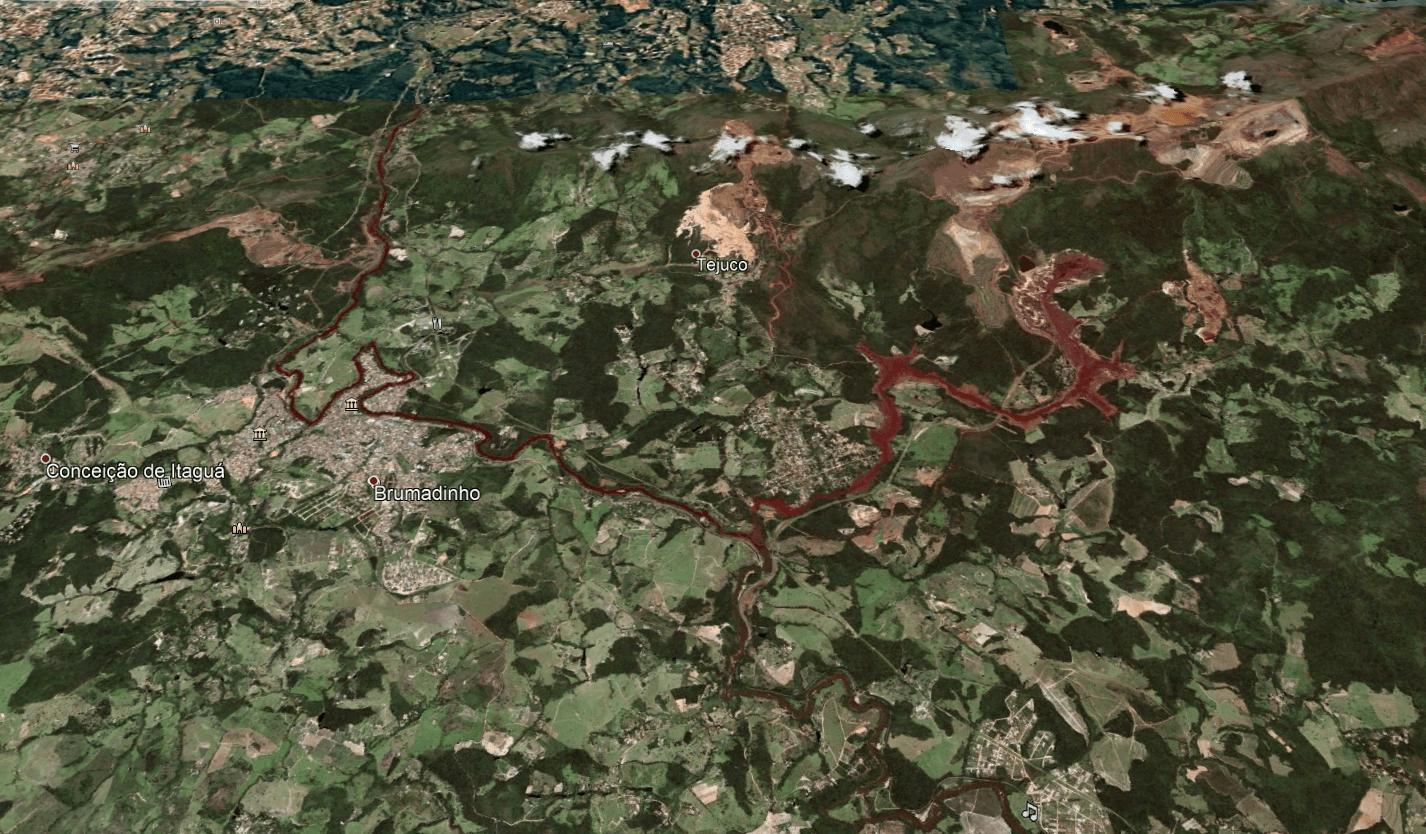}
      
      \caption{Extension of the mud-flow along the Paraopeba river}
      \label{fig:Jan2019}
    \end{minipage}
  \end{figure}

 The dam's destruction destroyed much of the city of Brumadinho as well as the surrounding landscape and infrastructure, including trails and freight railway tracks. The company Vale was responsible for the dam's failure due to neglect of safety regulations and was ordered to pay a fine of \$2.9bn~\cite{BBC19}. 
 Downstream from the dam is a narrow canyon leading to the river which was severely flooded by mud during the event as shown in Figure ~\ref{fig:Jan2019B} and ~\ref{fig:Jan2019}. The exact cause of the dam break has not been discovered. A complete analysis of its aftermath, total number of casualties and cost of property damage, has yet to be released. 

% \begin{figure}[!H]
%   \begin{minipage}[b]{0.45\linewidth}
%     \centering
%     
%     \includegraphics[width=9cm]{./1_25_c.jpg}
%     
%     \caption{An image of the dam on 25 January 2019 during collapse}
%     \label{fig:0125}
%   \end{minipage}
%   \hspace{1cm}
%   %
%   \begin{minipage}[b]{0.45\linewidth}
%     \centering
%     
%     \includegraphics[width=9cm]{./1_28_c.jpg}
%     
%     \caption{An image of the dam on 28 January 2019 during the settling of the mud}
%     \label{fig:0128}
%   \end{minipage}
% \end{figure}

%\begin{figure}[htp]
% \centering
% \includegraphics[width=4cm]{./2_1_c.jpg}
% \caption{An image of the dam on 1 February 2019 showing full aftermath of collapse}
% \label{fig:dam}
%\end{figure}

 \section{Theory of dam break modelling}
 The primary goal of this research is to study the flow of mud due to a tailings dam break and analyze its causes through rough modelling with flow analysis software (HEC-RAS)

 The fundamental 2D shallow water equations in flux form solved by HEC-RAS are 
 
 \begin{eqnarray} \label{sweq}
 \pdv{\mathbf{ U}}{t}+\pdv{\mathbf{ F_x}}{x}+\pdv{\mathbf{ F_y}}{y} &=& \mathbf {S} \\
 \mathbf{U} &=& \begin{bmatrix}
 h \\
 hu \\
 hv 
 \end{bmatrix} \\
 \mathbf{F_x} &=& \begin{bmatrix}
 hu \\
 huu + 1/2 g h^2 \\
 huv 
 \end{bmatrix} \\
 \mathbf{F_y} &=& \begin{bmatrix}
 hu \\
 huv \\
 hvv + 1/2 g h^2 \\
 \end{bmatrix} \\
 \mathbf{S} &=& \begin{bmatrix} 
 0 \\
 - g h \pdv{z}{x} - c_f u \\
 - g h \pdv{z}{y} - c_f v \\
 \end{bmatrix} 
 \end{eqnarray} 

 where $\mathbf{U}$ is the state vector containing height and momentum of fluid, $\mathbf{F_x}$ and $\mathbf{F_y}$ are the flux vectors
containing mass flux and momentum flux, $\mathbf{S}$ is the source term describing bottom topography and friction. The bottom friction coefficient $c_f$ can be calculated from the Manning formula $c_f = \frac{n^2 g \sqrt{u^2+v^2}}{R}$ and is dependent on the Manning's coefficient $n$ ~\cite{HRRM2016}. HEC-RAS also adds Coriolis terms to the source term when instructed.

\section{Modelling setup and result}

\begin{figure}[H]
    
      \centering
      
      \includegraphics[width=16cm]{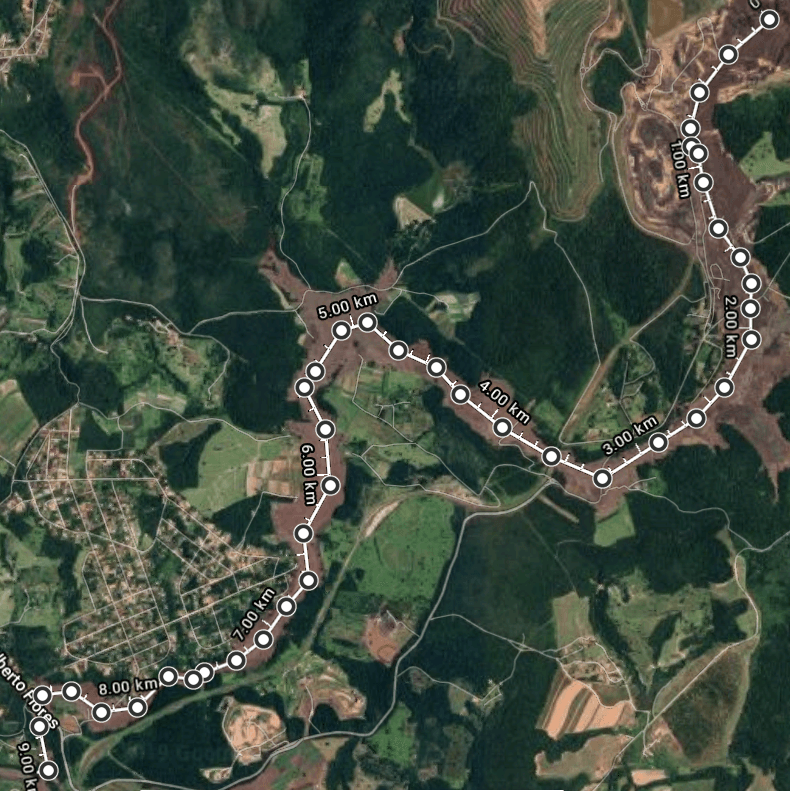}
      
      \caption{The measure distances of the flow path taken by the mud flow}
    
      \label{fig:floodextension}
  \end{figure}

The Digital Elevation Map (DEM) of the Brumadinho area was made available by NASA's Shuttle Radar Topography Mission (SRTM). This mission produced global 1 arc-second, or about 30 meters resolution, topographic data sufficient for the Brumadinho dam break simulation. As seen in Figure ~\ref{fig:floodextension}, the distance between the Brumadinho dam and Paraopeba river, the primary area where the flooding simulation is carried out, is approximately 9km which is the main area. The European Petroleum Survey Group (EPSG) code 29101 projection correctly geo-references the DEM to Google Satellite images over Brazil.

\begin{figure}[H]
    
      \centering
      
      \includegraphics[width=16cm]{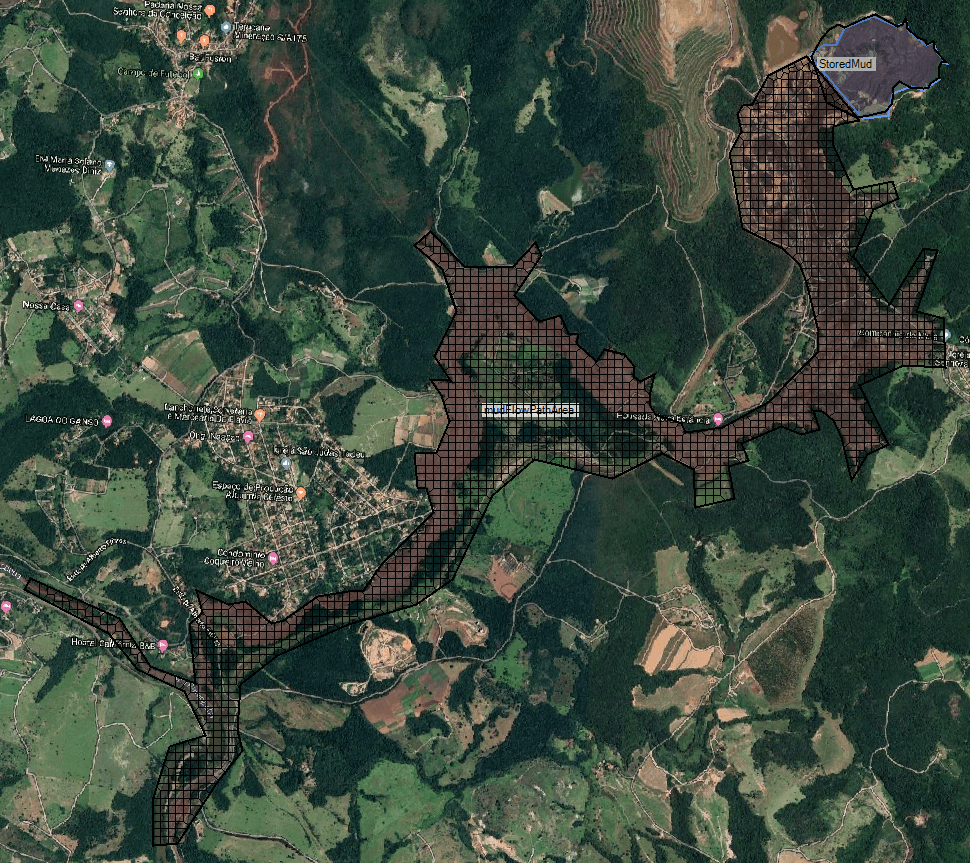}
      
      \caption{Computational mesh used in HEC-RAS for representing the assumed flow area of the simulation}
    
      \label{fig:cmesh}
  \end{figure}

A 2D flow area is set up with a 50 meter by 50 meter cell size resolution as show in Figure ~\ref{fig:cmesh}, resulting in a computational mesh with 2258 cells to carry out the shallow water equation calculations. A storage area is set up to store the mud with a 2D flow area and storage area connection in between to represent the dam.

\begin{figure}[H]
    \begin{minipage}[b]{0.9\linewidth}
      \centering
      
      \includegraphics[width=16cm]{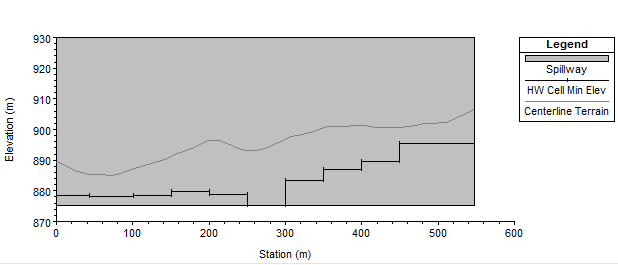}
      
      \caption{The dam modelled as 2D flow area and storage area connection}
    
      \label{fig:hras-thedam}
    \end{minipage}
  \end{figure}

The set up of the dam and breach plan is shown in Figure ~\ref{fig:hras-thedam}, which demonstrates that the left side of the dam is significantly lower than the right. This led to more significant erosion occurring to the left side of the breach during the dam break as can be seen in Figure ~\ref{fig:before_and_after}.

\begin{figure}[H]
    \begin{minipage}[b]{0.45\linewidth}
      \centering
      
      \includegraphics[width=9cm]{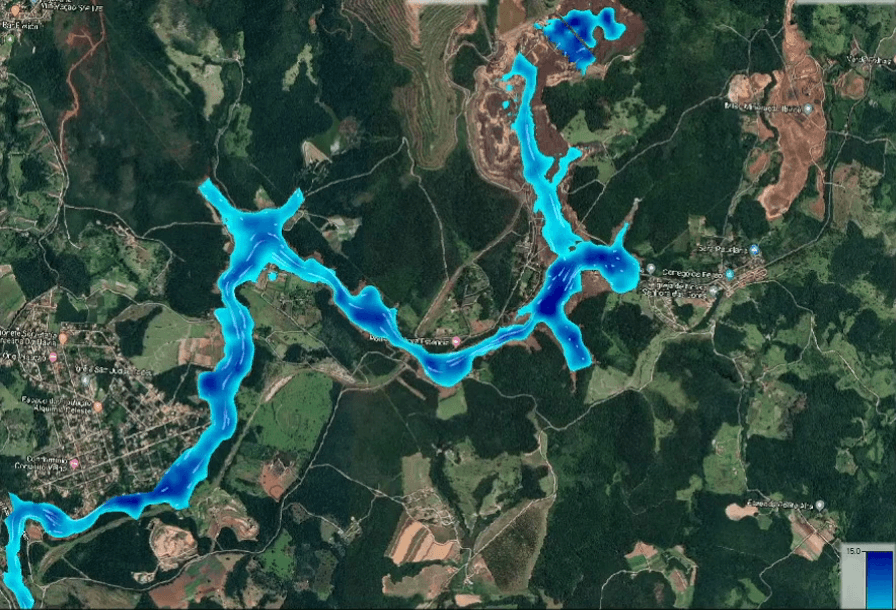}
      
      \caption{A simulation of the tailings flow area post-collapse}
      \label{fig:overall}
    \end{minipage}
    \hspace{1cm}
    \begin{minipage}[b]{0.45\linewidth}
      \centering
      
      \includegraphics[width=9cm]{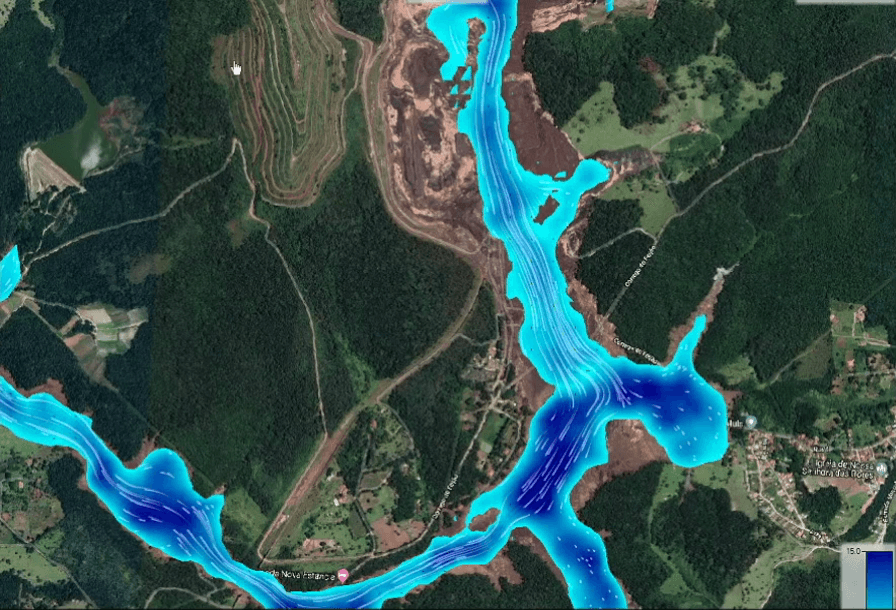}
      
      \caption{A closeup of the simulated tailings flow near the dam}
      \label{fig:closeup}
    \end{minipage}
  \end{figure}

The Manning value for the flow area is set to 0.3 which led to the observed 3 hour flow time from the Brumadinho dam to Paraopeba river. The Coriolis effect is turned off due to the relatively small scale (9 km) and short time (3 hours) involved in the simulation. Figure~\ref{fig:overall} and Figure~\ref{fig:closeup} show the result of the flooding pattern from the HEC-RAS simulation overlaid on top of the flooded region seen in the satellite image. The correlation between the observed flooding pattern and simulated pattern can be seen from the Figures to be extremely strong.

\begin{figure}[H]
    \begin{minipage}[b]{0.45\linewidth}
      \centering
      
      \includegraphics[width=9cm, height=6.83cm]{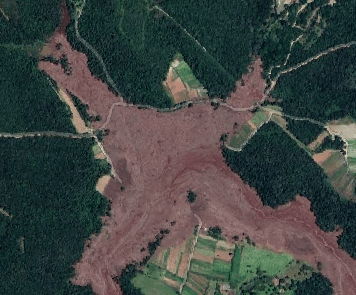}
      
      \caption{Loc. 1: Satellite Imagery}
      \label{fig:spot1_sat}
    \end{minipage}
    \hspace{1cm}
    \begin{minipage}[b]{0.45\linewidth}
      \centering
      
      \includegraphics[width=9cm]{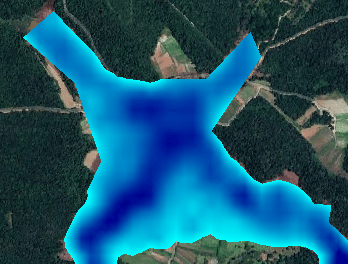}
      
      \caption{Loc. 1: HEC-RAS}
      \label{fig:spot1_sim}
    \end{minipage}
  \end{figure}

\begin{figure}[H]
    \begin{minipage}[b]{0.45\linewidth}
      \centering
      
      \includegraphics[width=9cm]{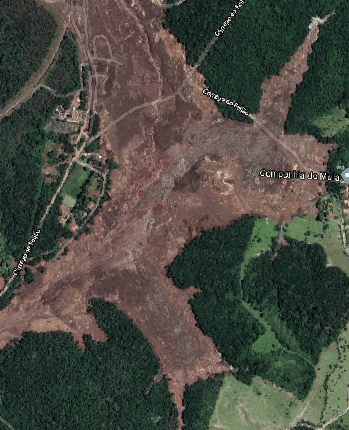}
      
      \caption{Loc. 2: Satellite Imagery}
      \label{fig:spot2_sat}
    \end{minipage}
    \hspace{1cm}
    \begin{minipage}[b]{0.45\linewidth}
      \centering
      
      \includegraphics[width=9cm, height=11.09cm]{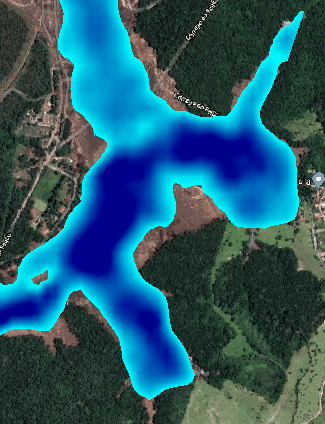}
      
      \caption{Loc. 2: HEC-RAS}
      \label{fig:spot2_sim}
    \end{minipage}
  \end{figure}

  A few isolated locations downstream from the broken dam are selected where we zoom in to examine the actual flooding and simulated pattern. These can be seen in Figure~\ref{fig:spot1_sat}, ~\ref{fig:spot1_sim}, ~\ref{fig:spot2_sat}, and ~\ref{fig:spot2_sim}. The images on the left are satellite images of the locations after the dam break; the images on the right are simulated flooding patterns. This level of similarity is seen in both the overall pattern comparison and at the individual locations examined.
  
  
\begin{figure}[H]
    \begin{minipage}[b]{0.45\linewidth}
      \centering
      
      \includegraphics[width=9cm,height=10cm]{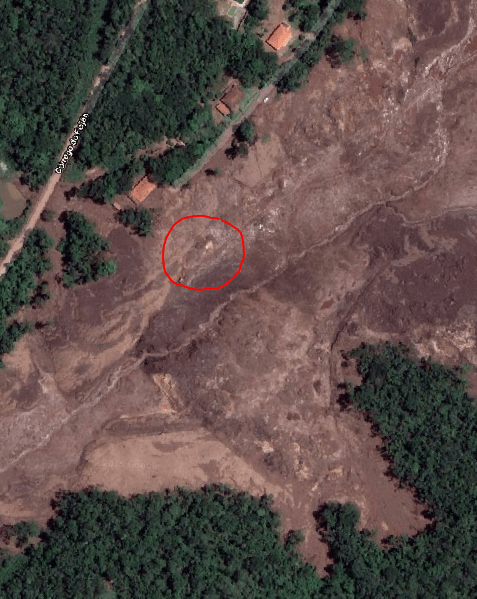}
      
      \caption{Loc. 3: Satellite Imagery}
      \label{fig:spot3_sat}
    \end{minipage}
    \hspace{1cm}
    \begin{minipage}[b]{0.45\linewidth}
      \centering
      
      \includegraphics[width=9cm,height=10cm]{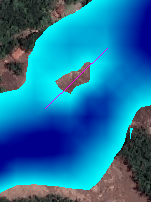}
      
      \caption{Loc. 3: HEC-RAS}
      \label{fig:spot3_sim}
    \end{minipage}
  \end{figure}

\begin{figure}[H]
      \centering
      \includegraphics[width=10.5cm]{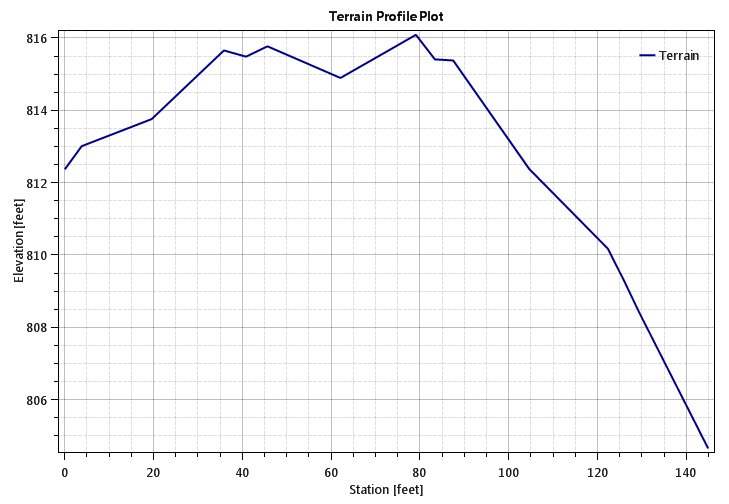}
      \caption{Terrain plot of Loc. 3 as shown above}
      \label{fig:spot3_terrain}
  \end{figure}

  A location of interest is shown in Figure~\ref{fig:spot3_sat} and ~\ref{fig:spot3_sim}. It can be seen that in the image of the simulated result ~\ref{fig:spot3_sim}, there is a small area without flooding. Upon closer examination of the satellite image and the bottom topography height along the line in ~\ref{fig:spot3_sim}, a local high ground is revealed ~\ref{fig:spot3_terrain} at this spot that pushes mud flow to circumvent around the local high ground.
  
% \begin{figure}[H]
%   \begin{minipage}[b]{0.45\linewidth}
%     \centering
%     
%     \includegraphics[width=9cm]{Spot4.PNG}
%     
%     \caption{Loc. 4: Satellite Imagery}
%     \label{fig:spot4_sat}
%   \end{minipage}
%   \hspace{1cm}
    %
%   \begin{minipage}[b]{0.45\linewidth}
%     \centering
%     
%     \includegraphics[width=9cm, height=10.62cm]{Spot4a.PNG}
%     
%     \caption{Loc. 4: HEC-RAS}
%     \label{fig:spot4_sim}
%   \end{minipage}
% \end{figure}
  
  \begin{figure}[H]
    \begin{minipage}[b]{0.45\linewidth}
      \centering
      
      \includegraphics[width=9cm, height=9.98cm]{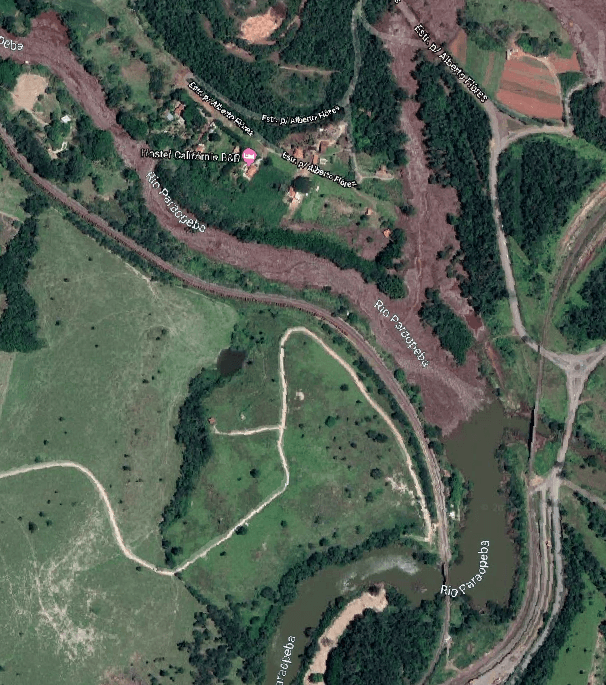}
      
      \caption{Location 5: Satellite Imagery}
      \label{fig:spot5_sat}
    \end{minipage}
    \hspace{1cm}
    \begin{minipage}[b]{0.45\linewidth}
      \centering
      
      \includegraphics[width=9cm]{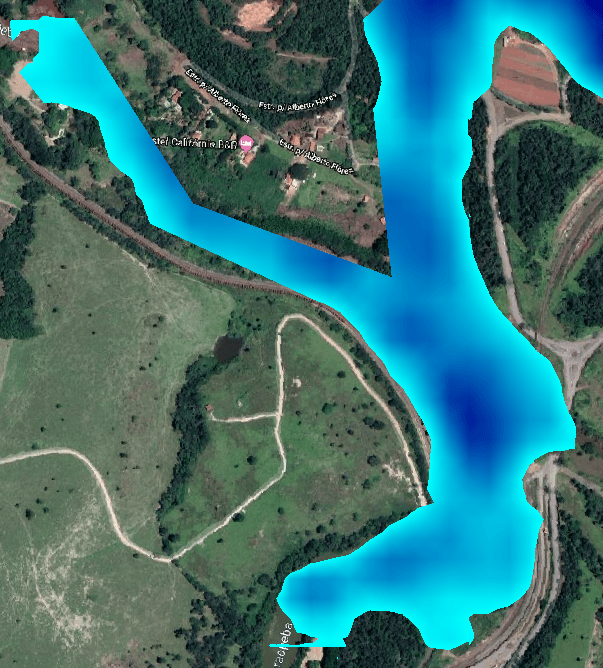}
      
      \caption{Loc. 5: HEC-RAS}
      \label{fig:spot5_sim}
    \end{minipage}
  \end{figure}
  
% images/vlcsnap-2019-09-22-00h57m05s179.png
%\begin{figure}[htp]
% \centering
% \includegraphics[width=4cm]{images/vlcsnap-2019-09-22-00h57m43s142.png}
% \caption{An image of the dam on 25 January 2019 during collapse}
% \label{fig:dam}
%\end{figure}
 
\section{Conclusions}
  Advancements in GIS technology have enabled large-scale simulations of dam break and flooding events through an interactive workflow using HEC-RAS and Google Earth. National and global satellite imaging has produced high-resolution topographic data available for creating bottom topography to be used in HEC-RAS for hydrology simulations. In this study, a digital elevation data set with a resolution of 1-arc second is made available by NASA's SRTM. Interested readers may want to know that over parts of the US, a topographic dataset with a resolution as high as 1/9 arc-seconds, or 4 m is available for use in simulations and studies. 
 
  The Brumadinho dam break, due to mishandling of mining waste and inadequate dam maintenance, caused significant human life and property loss. Such a tragedy could have been avoided if simple precautions were taken. For instance, it was erroneous to place the work lounge and cafeteria on the low-ground directly beneath the dam. The majority of the human casualties due to this event would have been avoided if the lounge and cafeteria were placed on the high ground on the right side of the dam. It can be seen from Figure~\ref{fig:before_and_after} that the right side of the dam was largely intact due to its elevation.
 
 A video captured during the onset of the dam break by a local TV station~\cite{YT2019} shows the bottom of the dam giving away due to structural failure leading to the entire dam's collapse on Jan 25, 2019. Over 12 million cubic meters of waste rushed downstream, contaminating the Paraopeba river with a large extension (Figure ~\ref{fig:floodextension}). The simulation of the dam break shows an astounding similarity between observed and modelled inundation patterns in overall and selected areas affected by Brumadinho break.
 
 While the aftermath of the Brumadinho dam break is still haunting us with many unresolved questions regarding the incident, a recent article from the WSJ ~\cite{WSJ2019} discusses the potential risks associated with building an upstream tailings dam for mining waste storage in Minnesota's Iron Range, where mining operations have lasted for over a century. According to the WSJ article, the exact cause of the dam break is unknown to the U.S. Mine Safety and Health Administration (MSHA), a federal agency that enforces compliance in mining with health and safety standards. As MSHA met with Brazil's National Mining Agency to discuss Brumadinho and deduce the exact cause of the dam break, PolyMet, the company responsible for the mining-waste disposal, insists that the failure of the upstream tailings dam at Brumadinho is no cause for a re-evaluation of the dam safety planned in Minnesota. As discussed earlier, upstream tailings dam designs are structurally unstable. This quantitative analysis of the potential damage due to Brumadinho dam break can help with the assessment of the mining dam construction in Minnesota, Brazil, and around the globe.

\end {document}